\documentclass[twocolumn, prl]{revtex4}
\usepackage{graphicx}
\usepackage{epsfig}
\begin{document}
\title{Criterion for bosonic superfluidity in an optical lattice}
\author{Roberto B. Diener, Qi Zhou, Hui Zhai, and Tin-Lun Ho}
\affiliation{Department of Physics, The Ohio State University, Columbus, OH 43210}
\date{\today}
\begin{abstract}
We show that the current method of determining superfluidity in optical lattices based on a visibly sharp bosonic momentum distribution $n({\bf k})$ can be misleading, for even a normal Bose gas can have a similarly sharp $n({\bf k})$. We show that superfluidity in a homogeneous system can be detected from the so-called visibility $(v)$ of $n({\bf k})$ $-$ that $v$ must be 1 within $O(N^{-2/3})$, where $N$ is the number of bosons.  We also show that the $T=0$ visibility of trapped lattice bosons is far higher than what is obtained in some current experiments, suggesting strong temperature effects and that these states can be normal.  These normal states allow one to explore the physics in the quantum critical region. 
\end{abstract}
\maketitle

There has been strong interest in using cold atoms in optical lattices to simulate strongly correlated 
many-body systems so as to shed light on many long standing problems in condensed matter physics. 
The interest began a few years ago with experiments on the superfluid-insulator transition of bosonic atoms in optical lattices~\cite{Bloch_Nature}, and has grown rapidly since the achievement of fermion pair condensation near a Feshbach resonance~\cite{Jin}. One class of very important problems, including high $T_{c}$ superconductivity, is understanding how superfluid order  (bosonic or fermionic) develops, and how the superfluid  transforms into other correlated many-body states as the interaction parameters are varied.  To achieve this goal, it is necessary to reach quantum degeneracy in a lattice, and to identify the presence of superfluidity. 

At present, the method commonly used for identifying superfluidity of bosons is through the ``sharpness" of the diffraction spots in their momentum distribution $n({\bf k})$
\cite{Bloch_Nature,Bloch_visibility_PRL,Esslinger,Sengstock,Ketterle}.  Despite its popularity, 
there has been no effort to characterize this ``sharpness" precisely.  As far as we can tell, a peak is considered ``sharp" if its width is visually much smaller than the separation of peaks.  
To be sure, a macroscopic bosonic superfluid is characterized by a $\delta$-function peak in $n({\bf k})$ (of order $N$ where $N$ is the number of particles). Unfortunately, the presence of such $\delta$-function is hard to discover due to finite experimental resolution. 
Instead, one relies on the 
estimate of sharpness mentioned above, 
which is consistent with but not a proof of superfluid correlation, as we explain below.

The purpose of this paper is to point out a number of facts crucial for identifying superfluid order for bosons in optical lattices. We show that 
{\bf (I)}  even a normal Bose gas above $T_{c}$ can have a diffraction pattern as sharp as those in current experiments. Identifying superfluid order from the sharpness of $n({\bf k})$ as practiced today is therefore unreliable. 
 {\bf (II)}  For homogeneous systems, the presence of superfluid order implies that the so-called ``visibility" $(v)$ must be 1 within $O(N^{-2/3})$. 
We also present {\bf (III)}  the visibility at $T=0$ as a function of lattice parameters for  the  ``wedding cake" structure of harmonically confined lattice bosons.  In this case, $v$ deviates from 1 when the superfluid regions are sufficiently small.  In current experiments, this typically occurs after {\it more than one} Mott layers have developed.  Because of the high sensitivity of $v$ to superfluid order, this visibility curve is a good calibration of temperature effects in the system.
These results have strong implications for the interpretation of many current experiments, discussed at the end. 

 
 In our discussions, we shall use the identification adopted in all current 
experiments~\cite{Bloch_Nature,Bloch_visibility_PRL,Esslinger,Sengstock,Ketterle} that 
 the observed diffraction pattern is related to the momentum distribution of the system through a ballistic expansion of the cloud (see eq.(\ref{nq}) below). While this has not been proven rigorously, it is consistent with the fact that the confinement energy of a Wannier state in the tight binding limit is much larger than the interaction energy~\cite{1D-expansion}.

{\bf (A) Normal Bose gas in a lattice:} Consider an ideal Bose gas with $N$ bosons in an optical  lattice with volume $\Omega$.  The Hamiltonian is  $H=\sum_{i}h_{i}$, $i=x,y,z$, $h_{i} = - (\hbar^2/2m)\partial_{i}^{2}+ V_{o}^{}{\rm sin}^{2}(\pi x_{i}/d)$, $V_{o}>0$.  
When $V_{o}$ is sufficiently large, only the lowest band (with energies $E_{\bf k} = -2t\sum_{i=x,y,z} \cos k_id $ and Bloch functions $\Psi_{\bf k}({\bf x})$) is thermally occupied. Above the Bose condensation temperature $T_{c}$, the chemical potential $\mu$ is determined by  
$N/\Omega = \Omega^{-1} \sum_{\bf k}'  \, f_{B}^{}(E_{\bf k})$, where  $f_{B}^{}(x) = (e^{(x-\mu)/k_{B}T} -1)^{-1}$, and 
 $ \sum_{\bf k}' $ is a sum over the first Brillouin Zone.  
At $T_{c}$, $\mu$ reaches the bottom of the band $E_{\bf 0}$.
The momentum distribution for  $T>T_{c}$ is 
$n({\bf q}) = \sum_{\bf k} ' \, f_{B}^{}(E_{\bf k}) |\tilde{\Psi}_{\bf k}({\bf q})|^2$,
where $\tilde{\Psi}_{\bf k}({\bf q})$ is the Fourier transform of $\Psi_{\bf k}({\bf x})$. 
Experimentally, one measures the {\em column} distribution, 
$N^{}_{\perp}({\bf q}_{\perp}^{}) =  \int {\rm d}q_{z}^{}  n({\bf q})$, where ${\bf q} = ({\bf q}_{\perp}^{}, q_{z}^{}) $.
   Since  $\tilde{\Psi}_{\bf k}({\bf q})$ is 
non-zero only when ${\bf q}= {\bf k} +{\bf G}$, where ${\bf G}$ is a reciprocal lattice vector, and since 
$E^{}_{\bf k}= E_{\bf k +G}^{}$, we have 
$N^{}_{\perp}({\bf q}_{\perp}^{}) =
 \int {\rm d}q_{z} f_{B}^{}(E^{}_{\bf q}) |\tilde{\Psi}_{\bf q-G}({\bf q})|^2 $, with ${\bf q-G}$ in the first Brillouin zone. 
 For a narrow band, $|\tilde{\Psi}_{\bf q-G}^{}({\bf q})|^2$ is a Gaussian centered at ${\bf q=0}$ decaying on the scale $2\pi/d$.
 Hence 
$N_{\perp}^{}({\bf q}_{\perp}^{})$ is composed of peaks centered at reciprocal lattice vectors $\bf G$, with the shape of the peak given entirely by the variation of $f^{}_{B}(E_{\bf q-G})$ around ${\bf G}$ and an overall envelope given by $|\tilde{\Psi}_{\bf q-G}^{}({\bf q})|^2$.

In Fig.~1(A) we show $N_{\perp}^{}({\bf q}_{\perp})$ for a lattice with one boson per site
($N/\Omega =d^{-3}$) and $V_{o} = 15E_{R}$ at a temperature $T= 1.1T_{c}$, 
where $E_{R}\equiv \hbar^2 \pi^2/(2md^2)$ is a ``recoil" energy.  
We have found numerically that in this system $k_{B}T_{c}= 0.45 B $, where $B =12t$ is the bandwidth of the lowest band, thus $k_B T < B$. 
As we shall see, this leads to sharp peaks distributed on a 2D 
square lattice with spacing $2\pi/d$.
Fig.~1(B) shows that the peaks are visibly ``sharp",  with a full width at half maximum $(\Delta q)_T  =0.1(2\pi/d)$.  This demonstrates that {\em diffraction spots in $n({\bf q})$ with width much less than a reciprocal lattice spacing is not proof of Bose condensation}.

\begin{figure}
\includegraphics[width=2.8in]{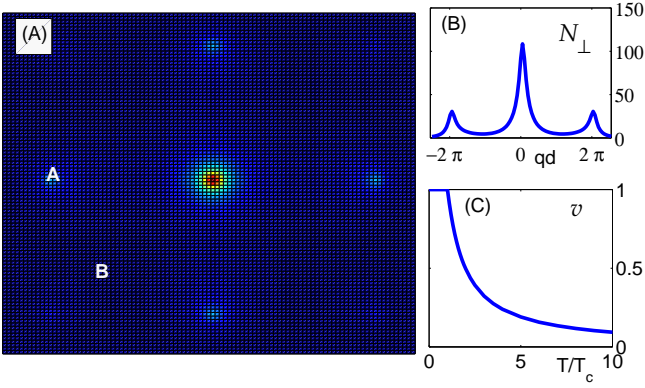}
\caption{(A): $N_{\perp}^{}({\bf q})$ for an ideal Bose gas with one particle per site in an optical lattice at  $T=1.1\, T_c$ for $V_0= 15 E_R$. $A$ and $B$ refer to those vectors defined in Eq.~(\ref{v}). 
(B):  $N_{\perp}^{}({\bf q})$ along the $q_{x}$-axis. (C): $v$ vs $T/T_c$ for this ideal Bose gas.}
\end{figure}

{\bf (B) Condition for quantum degeneracy:} 
The results in ${\bf (A)}$ can be understood by considering the condition for quantum degeneracy. 
When $k_{B}^{}T<B$, the most thermally occupied states are near the bottom of the energy band, for which we can use the approximate spectrum
$E_{\bf k} \approx -6t + \hbar^2 k^2/2m^{\ast}$, where $m^{\ast}$ is the effective mass defined as ($m^* = \hbar^2 /(2t d^2)$). The ``lattice" thermal wavelength  $\lambda_{T}= h/\sqrt{2\pi m^{\ast} k_{B}T}$ is reduced from the free space value $\lambda^{(o)}_{T}$ by $\sqrt{m/m^{\ast}}$.
The condition for Bose condensation for one boson per site, which is also the condition for quantum degeneracy ($\lambda_{T} \sim  d$),   becomes 
$k_{B}T_{c} = 0.55 B$.  The difference from the numerical result $(0.45B)$ is due to the effective mass approximation. The width of the spot is in general proportional to $\lambda_T^{-1}$.

Note that the change of thermal wavelength means that $T_{c}^{}$ in a lattice is reduced by a factor of 
$m/m^{\ast}$. This poses a severe challenge to reaching quantum degeneracy in the deep lattice limit.  For $V_{o}^{}/E_{R}^{} = 10, 15, 30$, we have $m/m^{\ast} = 0.25, 0.09, 0.007$ respectively. 
Without lattice, for gases with $10^{6}$ bosons in harmonic traps, $T_{c}$ is typically $10^{-6}$K and the lowest temperature attainable today  is $10^{-9}$K.  
For deep lattices with $m/m^{\ast}\sim 10^{-3}$,
one can barely reach quantum degeneracy even at the lowest temperature attainable today \cite{Tc}.

\begin{figure*}
$\begin{array}{c@{\hspace{0.2in}}c@{\hspace{0.2in}}c}
\epsfxsize=1.8in
 \epsffile{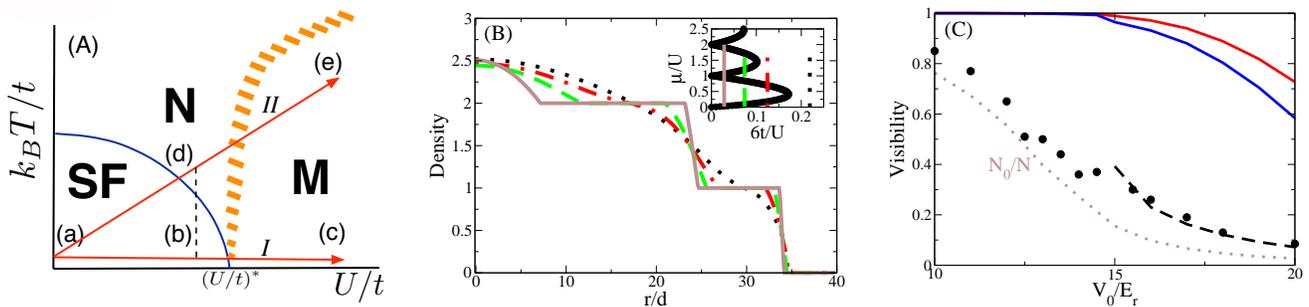}   &
\epsfxsize=2.2in
\epsffile{density_phase.eps}  &
        \epsfxsize=2.2in
\epsffile{visibility3.eps}    
        \end{array}$
\caption{(A):  Schematic phase diagram for a homogenous lattice Bose gas~\cite{indian}. The superfluid, (quantum critical) normal, and Mott insulator phases are labeled $SF$, $N$, and $M$ respectively.  The solid blue line is the critical temperature for the system.
The shaded area is a crossover.
$I$  and $II$ represent different physical processes. 
(B): Density vs radial distance for lattice bosons in a harmonic trap for conditions in ref.\cite{details of experiment} at $V_0 /E_R = 12$ (dotted black curve), 14 (dot-dashed red), 16 (dashed green), and 20 (solid brown).  Inset: phase diagram for a homogeneous system. 
The region outside the lobes is superfluid. Vertical lines represent values taken by $\mu({\bf r})$ for the corresponding $V_o$.   (C):  $T=0$ visibility vs $V_{0}$. For $V_0\le 14.7 E_{R}$, where the second Mott shell begins to appear, the system has a large superfluid core, which yields $v=1$. 
The upper solid red (lower, solid blue) curve is the visibility when the superfluids separated by the Mott shell are in (out of) phase.  The circles are experimental data from ref.\cite{Bloch visibility PRA}. The dashed line is a calculation including short range coherence assuming all superfluid regions have become Mott phases.  The dotted line is the superfluid fraction at $T=0$.}
\label{figtest-fig}
\end{figure*}

{\bf (C) Visibility and Bose condensation:}  The  
visibility, originally introduced to study short range coherence\cite{Bloch_visibility_PRL},  is defined as
\begin{equation}
v = \frac{N_{A} - N_{B}}{N_{A}+ N_{B}},
\label{v}  \end{equation}
where $N_{A} = N_{\perp}^{}(G\hat{\bf x})$, $N_{B} = N_{\perp}^{}(G\hat{\bf n})$, $G=2\pi/d$, $G\hat{\bf x}$ is a reciprocal lattice vector;  $G\hat{\bf n}$ is $G\hat{\bf x}$ rotated by $45^{o}$ around the $\hat{\bf z}$ axis and is not a reciprocal lattice vector.  
In the superfluid phase,  $N_{\perp}^{}(G\hat{\bf x})\sim N$ while $N_{\perp}^{}(G\hat{\bf n})\sim N^{1/3}$ (see later discussions), so we have  $v\approx 1$.  The visibility of an ideal Bose gas in a lattice with $V_{o}=15 E_{R}$ and one boson per site is shown in Fig.~1(C). 
The visibility is 100\% at $T<T_c$ but decreases sharply above $T_{c}$. It is interesting to note that despite its sharp drop at $T_{c}$, $v$ decays slowly, remaining at 0.1 at $T=10T_{c}$.  

For interacting bosons in a sufficiently deep lattice, the bosons are confined to the lowest band, described by the Bose Hubbard model 
$H = -t\sum_{\langle {\bf R, R'}\rangle} (a_{\bf R}^\dagger a_{\bf R'}^{} + h.c.) + {U\over 2} \sum_{\bf R} n_{\bf R}^{} (
n_{\bf R}^{}-1)$, 
where $\langle {\bf R, R'}\rangle$ means nearest neighbors, $a_{\bf R}^\dagger$ creates a boson in the Wannier state 
$w_{{\bf R}}({\bf r})= L^{-3/2}\sum_{\bf k}e^{-i{\bf k}\cdot {\bf R}}\Psi_{\bf k}({\bf r})$ 
located at site ${\bf R}$,
$L^3$ is the number of lattice sites, 
and $n_{\bf R} = a^{\dagger}_{\bf R}a^{}_{\bf R}$. The hopping integral $t$ and the interaction parameter $U$ are calculated from the eigenstates of $h_{i}$ and the s-wave scattering length in a straightforward manner.  Since the Fourier transform of $w_{\bf R}({\bf r})$ is of the form 
$w_{\bf R}^{}({\bf q}) = e^{-i{\bf q}\cdot {\bf R}} w({\bf q})$,  the momentum distribution is 
\begin{equation}
n({\bf q}) =   |w({\bf q})|^2 \sum_{\bf R,R'} \langle a^{\dagger}_{\bf R} a^{}_{\bf R'}\rangle 
e^{i {\bf q}\cdot ({\bf R}-{\bf R'})} . 
\label{nq} \end{equation}
In a homogeneous superfluid, $\langle a^{\dagger}_{\bf R} a^{}_{\bf R'}\rangle$ is essentially given by 
$|\Psi|^2$ for ${\bf R\neq R'}$, $\Psi  = \langle a_{\bf R} \rangle$\cite{shortrange}.  Denoting the number of condensed bosons as $N_{o}\equiv L^{3} |\Psi|^2$~\cite{depletion}, we have 
\begin{equation}
 n({\bf q})  =  \left[ (N-N_{o})  +   |\Psi|^2  f({\bf q} ) \right] |w({\bf q})|^2. 
\label{nqq}  \end{equation}
where $f({\bf q}) =|\sum_{\bf R}e^{i {\bf q}\cdot {\bf R}}|^2$.   
For a narrow band, $w({\bf q})$ is well approximated by  $|w({\bf q})|^2= \prod_{i=x,y,z}{\cal W}(q_{i})$, 
${\cal W}(k) = e^{-k^2/\sigma^2}/\sqrt{\pi \sigma^2}$, $\sigma \sim 1/d$. For a cubic lattice, we have  $f({\bf q})=\prod_{i}F(q_{i})$, $F(k) =  [{\rm sin}( Lkd/2)/ {\rm sin}(kd/2) ]^2$, which peaks sharply at reciprocal lattice vectors ${\bf G}$ with a width $\sim \pi/Ld$. 
Since the product $ {\cal W}(q_{x}){\cal W}(q_{y})$ has the same value at $G\hat{\bf x}$ and 
$G\hat{\bf n}$ and since $F(0)F(2\pi/d)= L^4$, it is simple to show from Eqs.~(\ref{v}) 
and (\ref{nqq}) that 
\begin{equation}
v = \frac{|\Psi|^2}{|\Psi|^2 + (N-N_{o})y},  \,\,\,\,\,
y = \frac{ 2\int {\cal W}(q_{z}) {\rm d}q_{z}}{  \int {\cal W}(q_{z})F(q_{z}) {\rm d}q_{z} L^4 } 
\end{equation} 
Simple integration shows that $y \approx  \frac{ d\, \sigma}{\sqrt{\pi}}\frac{1}{L^5} $. 
Since $|\Psi |^2 = N_{o}/L^3$, we then have {\em  in the superfluid phase}, 
$v = 1 -  O(1/N^{2/3})$\cite{Dimensionality}.
In order for the visibility to deviate from 1 by a non-zero but small amount, $v=1-\epsilon$, we need 
$N_{o}/N \sim 1/(\epsilon L^2) \sim 1/(\epsilon N^{2/3})$. Hence, even if the condensate fraction is very small, as long as it is larger than $1/(\epsilon N^{2/3})$, the visibility is essentially 1, ($1>v>1-\epsilon$ to be precise). For example, a visibility of 0.95 for a system with $10^{6}$ bosons with about a few bosons per site means $N_{o}/N \approx 10^{-3}$. 

{\bf (D) Visibility of lattice bosons in a harmonic trap:}   Many recent experiments investigating quantum phase transitions of a lattice Bose gas or Bose-Fermi mixtures have measured the visibility of these systems in harmonic traps as a function of lattice height $V_{o}$\cite{Bloch_visibility_PRL,Bloch visibility PRA,Esslinger,Sengstock}.  
Except for a single case in ref.\cite{Bloch_visibility_PRL} $(V_{o}=5E_{R}$)  which finds $v=1$ in the regime where a majority of bosons should be superfluid in the ground state, the data reported in ref.\cite{Bloch_visibility_PRL,Bloch visibility PRA,Esslinger,Sengstock} shows that  $v\leq 0.8$ in a similar regime. If there was no harmonic trap, a visibility $v=0.8$ means the lattice Bose gas is normal, as shown in Section ${\bf (C)}$.  On the other hand, in a harmonic potential $V({\bf r})$, it is well known that the system develops alternating layers of superfluid and Mott phases (the so-called ``wedding cake" structure).   When a sufficiently large number of bosons is converted from the superfluid phase to the Mott phase, the visibility will begin to drop. In addition, finite temperature or heating effects can also destroy phase coherence and reduce visibility.

To understand the general behavior of the visibility, 
let us consider a region in the harmonic trap  (say, around ${\bf r}$) where the lattice Bose gas turns into a Mott phase as the lattice depth $V_{o}$ (and hence the ratio $U/t$) increases. Within the local density approximation (LDA), we can treat this region as a bulk system 
for which the physical process is represented as a path in
the phase diagram of a homogenous lattice gas shown schematically in Fig.~2(A),  which plots  the transition temperature $T_{c}$ as a function of $U/t$. 
At $U/t=0$, 
 $T_{c}$ is given by the quantum degeneracy condition discussed in
Section ${\bf (B)}$. It drops to zero at the quantum critical point $(U/t)^{\ast}$.  The shaded line in Fig.~2(A) is the crossover from the quantum critical region (a normal phase with no clear sign of a gap) to the Mott region (a normal phase with an interaction gap).  

Since experiments are performed at finite temperature, any physical trajectory connecting the superfluid phase to the Mott phase must pass through the quantum critical region.  Typically, as $V_{o}$ increases, the system heats up due to a variety of reasons : spontaneous emission, tiny vibrations of the apparatus, etc. The physical processes may therefore look like trajectories $I$ or $II$ shown in Fig.~2(A). 
The states $(a)$ and $(b)$ are in the superfluid phase. The final states $(c)$ and $(e)$ are in the Mott regime.   The state $(d)$ is in the normal regime. 
The proximity to a quantum phase transition can be measured by the length of the trajectory passing through the quantum critical region.   For example, process $I$ is close to the quantum phase transition, whereas $II$ is not.   
In homogeneous systems, for both $I$ and $II$ one starts off with $v=1$ and a sharp momentum distribution $n({\bf k})$ in the superfluid region.  For $I$, $v$ drops sharply and $n({\bf k})$ becomes  blurry quickly across the transition point $(U/t)^{\ast}$.   For $II$, $v$ drops slowly as the system leaves the superfluid phase,  and  $n({\bf k})$ remains sharp in the quantum critical regime close to $T_{c}^{}$.   
 
 The proximity to a quantum phase transition can also be estimated by comparing the measured visibility to  the $T=0$ visibility calculated using standard mean field method\cite{indian} and  LDA. 
To be concrete, we focus on the system in ref.\cite{Bloch visibility PRA} because it has the most detailed analysis of data among  current experiments  on lattice bosons~\cite{details of experiment}. 
The physics illustrated here, however, should be applicable to boson-fermion 
mixtures\cite{Esslinger,Sengstock}.  
We begin by calculating the order parameter $\langle a^{}_{\bf R}\rangle = \Psi_{\bf R}$ 
 and density $\langle n_{\bf R}^{}\rangle $ of an infinite lattice as a  function of chemical potential $\mu$ and the interaction ratio $t/U$ for a homogeneous system.
The phase boundary between superfluid and Mott phases 
is a sequence of ``Mott lobes" as shown in the inset of Fig.~2(B).  The regions within different 
lobes are Mott phases with different (integer) number of bosons per site.
In a trap $V({\bf r})$, both  $\Psi_{\bf R}$ and $\langle n_{\bf R}^{} \rangle$ are position dependent, since $\mu$ becomes (within LDA) $\mu({\bf r})= \mu - V({\bf r})$.  
In this way, we obtain the density profiles in Fig.~2(B).  
 (We mention that our density profiles differ from those in ref.\cite{Bloch visibility PRA}, see \cite{wrong wedding cake structure}.  These differences, however, will not affect our points below.)
We then calculate $v$ from Eqs.~(\ref{v}), (\ref{nq}) with  $\langle a^{\dagger}_{\bf R} a^{}_{\bf R'}\rangle $ $=$ $\langle n_{\bf R}^{}\rangle \delta_{\bf R,R'}^{} + \Psi^{\ast}_{\bf R}\Psi^{}_{\bf R'}( 1 - \delta_{\bf R,R'}^{})$. 
If ${\bf R}$ and ${\bf R'}$ are in disconnected superfluid regions, the product $ \Psi^{\ast}_{\bf R}\Psi^{}_{\bf R'}$ depends on the relative phase $\Delta \theta$ between these regions. 

In Fig.~2(C), we have plotted our result for $v$ as a function of $V_{o}$ for the system in ref.\cite{Bloch visibility PRA}. 
The different curves correspond to different ways of treating the relative phase $\Delta \theta $ between different disconnected superfluid regions.
One sees that the $T=0$ visibility differs strongly from the experimental data (shown as circles in the figure). 
This large difference, however,  is {\rm not} due to the differences in our wedding cake 
structures. This is because both structures contain a superfluid core below $V_{0}=14.7 E_{R}^{}$ (where the second Mott shell begins to develop) that is so large that $v$ must be 1 as long as the superfluid is not destroyed. 
The disagreement with experiments {\em implies that all superfluid regions that should exist at $T=0$ have turned normal} (hence the much weaker visibility),  which can only occur if  the temperature is above  $T_{c}$ in these regions (i.e. the system that should be in the 
superfluid state $(b)$ in Fig.~2(A) at $T=0$  is found to be in state $(d)$ above $T_{c}$).   
The physical process is therefore quite far from the quantum critical trajectory~\cite{Bloch_heating}.

In order to account for the visibility deep in the Mott regime, 
the authors of ref.\cite{Bloch visibility PRA} 
considered short range correlations in a perturbative manner and found good agreement with their data,  {\em provided one makes the assumption} that all the superfluid regions are converted into the Mott phase.  We have repeated this procedure with our wedding cake structure~\cite{wrong wedding cake structure} and have obtained similar agreement (dashed line in Fig.~2(C)). 
The assumption that lead to this agreement, which eliminates all contributions from superfluid to visibility, is consistent with the picture that all superfluid regions has gone normal due to temperature effects. 
We would like to point out that our mean field calculations do not include these short range coherence but that their inclusion would only raise the visibility curves in Fig.~2(C) to even larger values.
Finally, it is instructive to look at the condensate fraction $N_{0}/N$ at $T=0$ as a  function of $V_{0}$, where $N_{0} = \sum_{\bf R}^{} |\Psi_{\bf R}|^2$.  We see from Fig.~2(C) that at $T=0$ a visibility as high as 0.8 (which occurs when $V_{o}>14.7E_{R}$) represents a condensate fraction $N_{0}/N \sim 0.05$.

{\bf (E) Implications for recent experiments:} 
Just as in ref.\cite{Bloch visibility PRA}, ref.\cite{Esslinger,Sengstock} also show visibility $v\sim 0.8$ for lattice heights where the system should be superfluid at $T=0$.  
The physical processes in ref.\cite{Esslinger,Sengstock} are therefore quite far from being a quantum phase transition (QPT). 
(To show that a physical trajectory  is close to a QPT, it is necessary to demonstrate that the quantum critical region traversed in the process is very narrow).  
Moreover, the decrease of visibility of the bosons when fermions are added suggests that fermions may be increasing the temperature of bosons. 
Our discussions in  Section ${\bf (A)}$ and ${\bf (B)}$ also show that 
the recent claim of observation of superfluid correlation of fermions in an optical lattice\cite{Ketterle}  based on the sharpness of $n({\bf k})$ 
is not conclusive. The claim would have been established if the bosonic molecules after the sweep were found to have visibility $v=1$. 

Our study indicates that the problem of heating is prevalent in current experiments. We hope our findings will stimulate serious efforts to determine the temperature of lattice gases and more rigorous ways to achieve quantum degeneracy in lattices. 
This work is supported by NSF Grant DMR-0426149 and PHY-0555576.


\begin{thebibliography}{99}
\bibitem{Tc}Although for a non-interacting gas the ratio $T/t$ remains constant as the lattice is adiabatically turned on, once interactions are included this ceases to be the case.
\bibitem{Dimensionality}
The exponent $2/3$ is valid in 3D due to the integration along $q_z$.  In 1 and 2D, $v \sim 1- O(1/N)$.
\bibitem{details of experiment}
Ref.\cite{Bloch visibility PRA} has $N=2.2 \times 10^5$, lattice period $d = 425$ nm,  trap frequency $\Omega = \sqrt{\omega_m^2 + 8 V_0/(mw^2)}$, where $\omega_m = 2\pi \times 15$ Hz, $w = 130 \mu$m, and $m$ is the atomic mass,
\bibitem{wrong wedding cake structure} In Fig. 1 of ref.\cite{Bloch visibility PRA}, there is no superfluid core at $V_0 = 18 E_R$. Within the same mean field and LDA, we find a superfluid core even for larger values of $V_{0}^{}$. 
We have tried to verify the cake structure in ref.\cite{Bloch visibility PRA} using their formulae for $t$ and $U$ but have recovered our structure instead.  We also note that the results in ref.\cite{Bloch visibility PRA} are inconsistent with the parameters therein. 
The radii ($R_1$ and $R_2$) of the Mott layers are related to the interaction $U$ as $U = \Delta E$, $\Delta E={1\over 2} m \Omega^2 (R_2^2 - R_1^2)$. Using the results in  ref.\cite{details of experiment},  
$R_1 = 35 d$, $R_2 = 22 d$, and their 
$\Omega$ in (see footnote \cite{details of experiment}), we have $\Delta E = 0.62 E_{R}$. 
This is $30\%$ larger than the value obtained using their formula 
$U/E_R = 5.97 (a/\lambda_L)(V_0/E_R)^{0.88}$, where  $a = 5.45$ nm is the scattering length 
of $^{87}$Rb.
\end{thebibliography}
\end{document}